\documentclass[twocolumn,twoside,preprintnumbers,supercriptaddress,amsmath,amssymb]{revtex4}
\usepackage{epsfig}
\usepackage{graphicx}
\usepackage{dcolumn}
\usepackage{bm}
\usepackage{fancyhdr}
\usepackage{pslatex}
\begin{document}

\title{\Large First results from the Pierre Auger Observatory}

\author{Ronald Cintra Shellard for the Pierre Auger Collaboration}
\affiliation{Centro Brasileiro de Pesquisas F\'{\i}sicas\\Rua Dr.
Xavier Sigaud 150, Rio de Janeiro, RJ, 22290-180, Brazil}

\received{\today} 

\begin{abstract}We review in these notes the status of the
construction of the Pierre Auger Observatory and present the first
Physics results, based on the data collected during the first year
and a half of operation. These results are preliminary, once the
work to understand the systematics of the detectors are still
underway. We discuss the cosmic ray spectrum above 3 EeV, based on
the measurement done using the Surface Detector and the Fluorescence
Detector, both, components of the observatory.  We discuss, as well,
the search for anisotropy near the Galactic Center and the limit on
the photon fraction at the highest energies. \end{abstract}

%

\bibliographystyle{unsrt}

\maketitle
\thispagestyle{fancy}

\section{Introduction}

The cosmic ray spectrum, measured at the top of the atmosphere
covers a huge range in energy, going from 10~MeV to energies above
$10^{20}$ eV, with a differential flux that spans 31 decades
\cite{nag00}. The techniques to survey this spectrum goes from
instruments aboard satellite flights, balloon borne detectors, to
counters that monitor the fluxes of neutrons and muons at the Earth
surface, and at higher energies to wide area arrays of particle
detectors. The spectrum can be divided roughly into four regions
with very distinct behavior. The first one, with energies below
1~GeV, has a very distinctive character from the rest of it. Its
shape and cut-off is strongly dependent on the phase of the solar
cycle, a phenomenon known as {\it solar modulation}. Actually, there
is an inverse correlation between the intensity of cosmic rays at
the top of the atmosphere and the level of the solar activity
\cite{lon92,she85}.

The region above 1~GeV show a spectrum with a power law
dependence, $N(E)$~d$E=K~E^{-x}$~d$E$, where the spectral index
$x$ varies as $2.7~<~x~<~3.2$. The region between 1~GeV and the
{\it knee} region at $4\times10^{15}$~eV,  is characterized by an
index $x~\simeq~2.7$. These cosmic rays most likely are produced
at supernova explosions and their remnants  \cite{aha04}.  At
the knee ($4\times10^{15}$~eV) the power law index steepens to 3.2
until the so called {\em ankle}, at $5\times10^{18}$~eV. The
origin of the cosmic rays in this region is less clear and subject
of much conjecture. Above the ankle the spectrum flattens again to
an index $x~\simeq~2.8$ and this is interpreted by many authors as
a cross over from the steeper galactic component to a harder extra
galactic source for the cosmic rays  \cite{cro99,oli00}.

The existence of cosmic rays with energies above $10^{20}$~eV
presents a puzzling problem in high energy astrophysics. The first
event of this class of phenomena was observed at the beginning of
the 60's by John Linsley at the Volcano Ranch experiment, in New
Mexico  \cite{lin63,lin63a}. Since then, many events were
detected, at different sites, using quite distinct techniques
\cite{law91,efi91,bir93,yos97,tak98}. If those cosmic rays are
common matter, that is, protons, nuclei or even photons, they
undergo  well known nuclear and electromagnetic processes, during
their propagation through space. Their energies are degraded by the
interaction with the cosmic radiation background ({\sc cmb}), way
before they reached the Earth  \cite{gre66,zat66}. After
travelling distances at the scale of 50~Mpc, their energies should
be under $10^{20}$~eV, thus restricting the possible conventional
astrophysical objects, which could be sources of them, and those
should be easily localized through astronomical instruments. On the
other hand, it is very difficult to explain the acceleration of
charged particles, with energies up to $3~\times~10^{20}$~eV or even
greater, on known astrophysical objects, by means of electromagnetic
forces, the only conventional ones capable of long range and long
periods of acceleration.

\section{The Pierre Auger Observatory}

The Pierre Auger Observatory  \cite{aug96,aug01} was designed to
study the higher -- above $10^{18}$ eV -- end of the cosmic ray
spectrum, with high statistics over the whole sky. The detectors are
optimized to measure the energy spectrum, the directions of arrival
and the chemical composition of the cosmic rays, using two
complementary techniques, surface detectors (SD) based on Cherenkov
radiators and fluorescence light detectors (FD). The complete
project calls for two sites, one in the southern hemisphere, well
ahead in its construction, and a northern site, in order to achive
the homogeneous coverage of the whole sky, essential to pinpoint the
sources of the ultra high-energy cosmic rays.

\begin{figure}[htbp]
\centerline{\includegraphics[width=7.7cm]{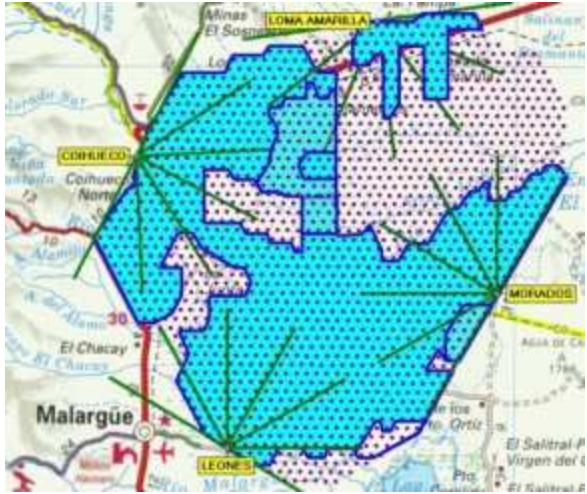}}
\caption{Layout of the Pierre Auger Observatory southern site
showing the four FD telescopes ({\em eyes}). The dots represents the
deployement location of the SD stations. The darker shade represents
the area instrumented in July 2006. The FD eyes Los Leones, Los
Morados and Coihueco are in full operation, while the construction
of Loma Amarilla and its instrumentation should be completed by the
end of 2006.}
\label{rcsfig01}
\end{figure}

The southern site of the Observatory, located in the province of
Mendoza, in Argentina, at the latitude 35$^\circ$ South and
longitude 69$^\circ$ West, on a very flat plateau at 1~400~m above
sea level, is bounded by the Andes Cordillera on the west. The
region has a very clear sky, with little light pollution, essential
for the operation of the fluorescence system. The main office of the
observatory is located at the northern entrance of the City of
Malarg\"{u}e. The construction of this part of the detector should
be completed on the second half of 2007. The layout of the site is
shown in Figure \ref{rcsfig01}, indicating the distribution of
detectors already deployed and the position of the fluorescence
buildings.

The construction of the northern site, in the southern part of the
state of Colorado, at the United States, will start after the full
completion of the southern site.

The initial phase of the Pierre Auger Observatory, the {\em
Engineering Array} was built during the period 2000-2001. Then, 40
Cherenkov tanks were laid and instrumented, to test the components
of the array and to prove the soundness of its design. They were set
in a roughly hexagonal array, covering an area of 54~km$^2$. Two of
the tanks, at the center of the array, were laid side by side, in
order to cross calibrate their signals. Two prototype telescopes
were installed on the Los Leones hill, overlooking this ground
array. They were successfully tested during the southern summer of
2001-2002. The lessons learned in the Engineering Array lead to
improvement in the final design of the detector components
\cite{aug04}.

\subsection{The Surface Detector}

The Surface Detector (SD) is a ground array which spans an area of
3~000~km$^2$, with 1~600 stations set on a regular triangular grid,
with 1~500~m separation between them. The stations communicate with
the central base station through a radio link.

Each station at the SD is a cylindrical tank,  filled with 12~000~l
of purified water, operating as a Cherenkov light detector. The
tanks are manufactured by rotational molding process, using a high
density polyethylene resin, with 12.7 mm thick walls, opaque to
external light. The walls have two layers, an external in the color
beige, to minimize the environmental impact and reduce the heat
absorption, and the internal black, with the addition of carbon
black to the resin. The water is contained in a liner inside the
tank, a bag made of a sandwich of polyolefin-Tyvek film. The Tyvek
film has a high reflectivity to ultraviolet light and its role is to
diffuse the UV Cherenkov light within the volume of water.

The tanks are prepared in the Assembly Building, just behid the Main
campus building. The water purification plant produces enough  water
for 3 tanks per day, with a resistivity of 15~M$\Omega$ per cm. The
tanks are manufactured in S\~{a}o Paulo, Brazil, and Buenos Aires
and shipped all the way to Malarg\"{u}e on flatbed trucks, carrying
six tanks on a load. After preparation which includes the mounting
of the electronic components, the cabling and fitting the internal
Tyvek bag, the tanks are deployed on the field, on prepared ground,
using 4-wheel drive vehicles.

\begin{figure}[htbp]
\centerline{\includegraphics[width=9.4cm]{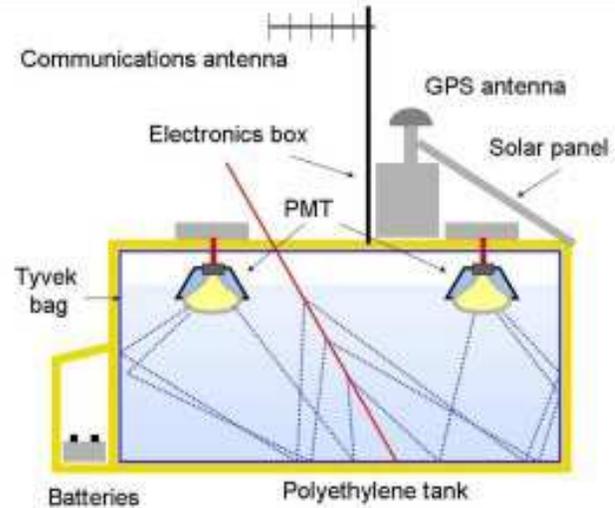}}
\caption{A schematic view of the Cherenkov water tanks, with the
components indicated in the figure.}
\label{rcsfig02}
\end{figure}

The Cherenkov light, diffused within the volume of water, is
collected by three 20.3 cm diameter photomultiplier tubes (Photonis
XP1805), set in a symmetric pattern on top of the tanks, facing
downwards. This arrangement avoids the direct hit of the Cherenkov
light, collecting a signal which is homogeneous and proportional to
the length of charged tracks crossing the water. A schematic view of
a SD station is shown in Figure \ref{rcsfig02}.

The electronics for the detectors are housed in a box outside the
tanks and communicate with a base station through a radio {\sc
wlan}, operating in the 915~MHz band. The stations are powered by
a bank of two special 12 V batteries,  which are fed by two solar
panels of 55~W each. The time synchronization of the tanks is
based on a {\sc gps} system, capable of a time alignment precision
of about 10~ns  \cite{pri95}. A 7 GHz microwave backbone links
the base stations to the central data acquisition station (CDAS),
at the campus of Malarg\"{u}e in the southern site.

Each detector station has a two level trigger, a hardware
implemented T1 and a software T2. The T1 trigger is decided in a PLD
(Programmable Logical Device) \cite{Allard:2005vk}, set with a
threshold defined in terms of a vertical equivalent muon (VEM)
crossing a tank, with a typical value  of 1.75 VEM on a single 25 ns
time bin, in coincidence at all working PMT's. This trigger has a
rate of about 100 Hz and is necessary to detect fast signals
associated to the muons of horizontal showers. There is a second
condition for T1, which is a time over threshold condition (ToT),
requiring that the signal in 13 FADC bins out of a window of 120
bins are above a value of 0.2 VEM. The ToT trigger rate averages 1.6
Hz, dominated by coincidences of two muons and is efficient to
select small signals away from the core of showers at the tail of
the lateral distribution. T1 is adjusted so that its rate is about
100~Hz. T2 limits the  trigger rate at each station to less than
20~Hz, so as not to saturate the radio bandwidth available. ToT
triggers are promptly promoted to T2, while the threshold T1 are
requested a tighter condition of 3.2 VEM in coincidence for the 3
PMTs. The event trigger (T3) is set at CDAS combining the triggers
of contiguous individual stations. The requirement in the number of
stations triggered sets the lower energy threshold. Typically, four
stations are required for a threshold of 10$^{19}$ eV. The
communication between the central campus and the individual stations
is bi-directional to allow the T3 trigger request the data to be
downloaded to the CDAS system.

\begin{figure}[htbp]
\centerline{\includegraphics[width=8.9cm]{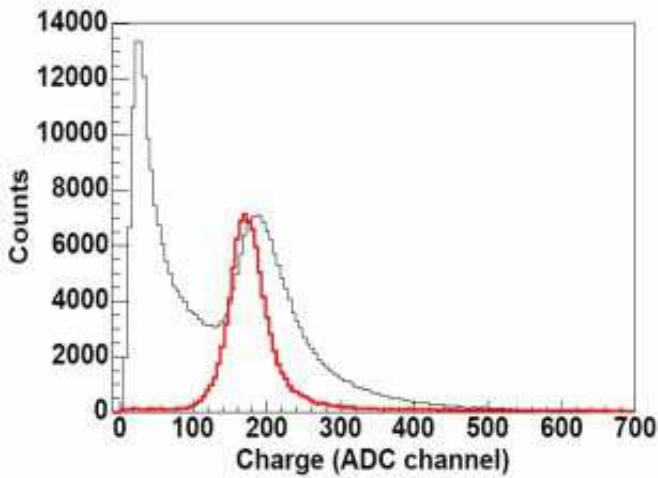}}
\caption{Typical histogram used in the calibration of the surface
stations. This was taken from a prototype tank equipped with
scintillation pads to select the component of the signal due to
vertical muons. The hump of the experimental distribution of single
muons is related to the signal due to a vertical incoming muon
\cite{Aglietta:2005xx}.}
\label{rcsfig03}
\end{figure}

The calibration of the surface stations is done continuously. At
every minute the histogram of low energy particles is taken,
correspondinding roughly to about 100~000 events, mostly atmospheric
muons coming from all angles into the detector. This calibration is
performed in parallel to the data taking and at every 6 minutes this
is sent to the central data acquisision system (CDAS) for
monitoring \cite{Aglietta:2005xx}. Figure \ref{rcsfig03} shows a the
histogram for a surface station, taken at a prototype detector.

The incoming angle of a shower is reconstructed from the timing of
arrival of the signals in the tanks. The detectors have a large
cross section even for a shower with very high zenith angle, so that
Auger is quite sensitive to neutrinos \cite{cap98, ber02}.  A
highly inclined shower, originating from a hadronic cosmic ray, has
a very characteristic signature, having lost a substantial part of
its electromagnetic component, with only a core of energetic muons
remaining. The muons arrive concentrated in time, generating signals
with a very sharp peak. In contrast, a shower which has a large
electromagnetic component is spread in time, with a much more
complex structure.

\subsection{The Fluorescence Detector}

The Fluorescence Detector (FD) is composed of 4 eyes disposed on
the vertices of a diamond-like configuration,  with a separation
of 65.7~km along the axis South-North (actually this axis is
tilted by 19$^\circ$ into Northeast) and 57.0~km in the West-East
direction (tilted by 20$^\circ$ into Southeast). They are at the
periphery of the ground array and all stations of the SD are contained
in the field of view (FOV) of the FD telescopes \cite{Bellido:2005yq} (see
Figure \ref{rcsfig01}).

\begin{figure}[htbp]
\centering\includegraphics[width=8.3cm]{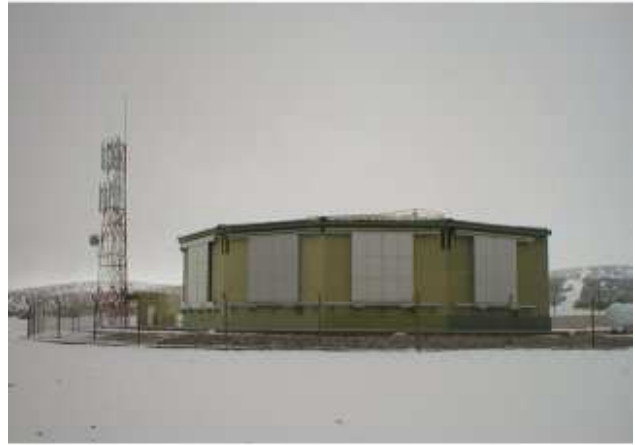}
\centering\includegraphics[width=8.2cm]{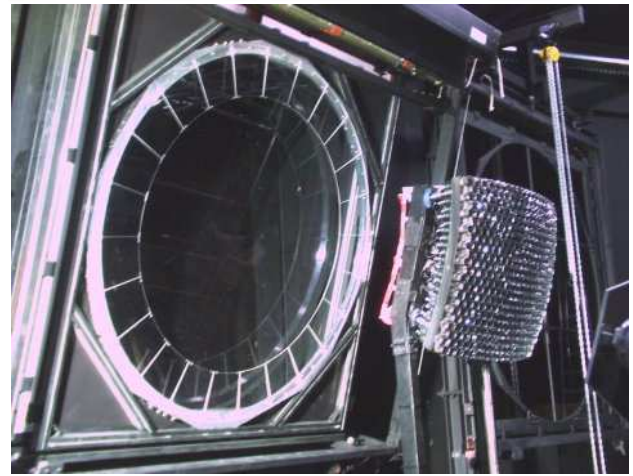}
\caption{The Coihueco telescope building, with the shutters closed.
Lower picture: The diaphragm of a telescope, with corrector rings. The
picture shows the camera, with its 440 pixels facing the mirror.}
\label{rcsfig04}
\end{figure}

Each eye has six independent telescopes, each with a field of view
of $30^\circ$ in azimuth and $28.6^\circ$ in elevation, adding to a
$180^\circ$ view of the array. Figure \ref{rcsfig04} shows the
building of the Coihueco eye located on the west side of the array
(see Figure \ref{rcsfig01}). The fluorescence light is collected by a
mirror with a radius of 3.4~m and reflected into a camera, located
at the focal surface of the mirror. The telescopes use a Schmidt
optics design to avoid coma aberration, with a diaphragm, at the
center of curvature of the mirror, with an external radius of
0.85~m. The advantage of this scheme is to project a point of light
in the sky into a reasonably homogeneous spot, at the focal surface,
of 0.25$^\circ$ radius.

The shape of the telescope mirror is a square with rounded corners
with a side of 3.8~m. The radius of the diaphragm may be enlarged to
1.1~m, by adding a corrector lens annulus on the area with radius
between 0.85~m and 1.1~m  \cite{Sato:2005jb} (see Figure
\ref{rcsfig04}). The telescope, which is remotely operated, is
protected from the environment by an external shutter and a MUG-6 UV
transmitting filter, which reduces the light background.

The camera is composed by 440 pixels, each a hexagonal
photomultiplier (Photonis XP3062), which monitors a solid angle of
(1.5$^\circ$)$^2$ projected onto the sky. The dead spaces between
the photomultipliers is corrected by a reflective wedge device,
called the {\em Mercedes corrector}, making the sky exposure very
uniform.

The electronics of the FD detector was designed to be operated
remotely and with flexibility to reprogram the trigger to
accommodate non-standard physical processes which may show up. The
trigger rate is not limited by the transmission band available. The
FD communicates with the CDAS system through the 7 GHz microwave
backbone, with a capacity of  34 Mbps throughput.

To be able to operate in hybrid mode the absolute time alignment
with the Cherenkov water detector must be better than 120 ns. The
trigger for each FD pixel is programmed to have a rate below 100~Hz,
while a special processor, with  a built-in pattern recognition
algorithm, keeps the overall trigger rate below 0.1~Hz.  The pixels
are sampled and the signal digitized at every 100~ns. The first
level trigger uses a boxcar type addition of the signal over a 1
$\mu$s bin and a pixel trigger set whenever this sum is 3 $\sigma$
above the background noise. An event is triggered whenever a set of
5 contiguous pixels are triggered with a time sequence associated
\cite{kle04}.

The calibration of the camera is performed by exposing the PMTs to
signals from calibrated light sources diffused  from an aparatus
mounted on the external window of the telescopes. The camera is
illuminated uniformly and the gain of each PMT characterized
\cite{Aramo:2005jk, Bauleo:2005za}.

The determination of the shower energy requires an accurate estimate
of the atmosphere attenuation, due to the Rayleigh (molecular) and
Mie (aerosol) scattering,  of the light emitted by the fluorescence
excitation of the atmosphere. This is done by a complex set of
detectors, the Horizontal Attenuation Monitor (HAM), the Aerosol
Phase Function monitors (APF) and Lidar systems
\cite{mus04,Cester:2005hk} mounted at each eye. The system is
complemented by sky monitoring CCD's, which measures the light
output of selected stars. A set of infrared cameras at each eye
monitors the cloud coverage at the site. Monitoring the background
noise due to stars in the field of view of the pixels are tools used
to get the proper alignment of the telescopes.

\begin{figure}[htb]
\centering\includegraphics[width=6.5cm]{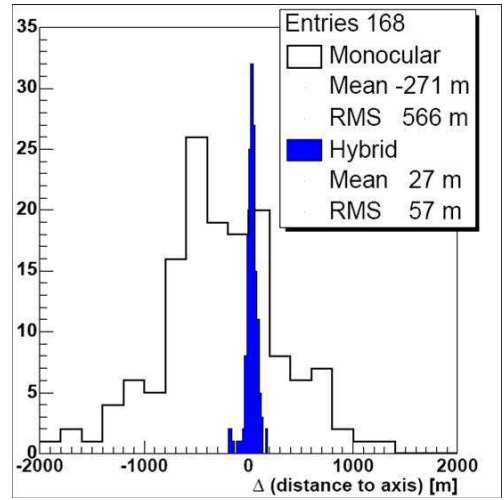}
\caption{Resolution of the reconstructed position for vertical
laser shots from the CLF.}
\label{rcsfig05}
\end{figure}

The Central Laser Facility (CLF) \cite{Arqueros:2005yn} is a
steerable automatic system which produces regular pulses of linearly
polarized UV light at 355nm. It is located in the middle of the
array, 26 km away from Los Leones and 34 km from the Coiheco eye.
This system provides a complementary measurement of the aerosol
vertical optical depth {\it versus} height, and the horizontal
uniformity  of the atmosphere across the aperture of the array. This
system creates an artificial cosmic ray by feeding a signal into a
nearby  tank through a fiber optics cable. The laser track in the
atmosphere is intense enough to be registred by all fluorescence
detectors. The time recorded at each detector can be used to measure
and monitor the relative timing between the SD stations and the FD
eyes. This time offset has been measured to better than 50 ns
\cite{Allison:2005vj}, leading to a systematic uncertainty in the
core location of 20 m. This is shown in Figure \ref{rcsfig05}.

Due to the much improved angular accuracy, the hybrid data sample is
ideal for anisotropy studies. Many ground parameters, like the shower
front curvature and thickness, have always been difficult to measure
experimentally, and were usually determined from Monte Carlos
simulation. The hybrid sample provides a unique opportunity in this
respect. As mentioned, the geometrical reconstruction can be done
using only one ground station, thus all the remaining detectors can
be used to measure the shower characteristics. The possibility of
studying the same set of air showers with two independent methods is
valuable in understanding the strengths and limitations of each
technique. The hybrid analysis benefits from the calorimetry of the
fluorescence technique and the uniformity of the surface detector
aperture \cite{Mostafa:2005kd}.

\begin{figure}[htb]
\centering\includegraphics[width=8.4cm]{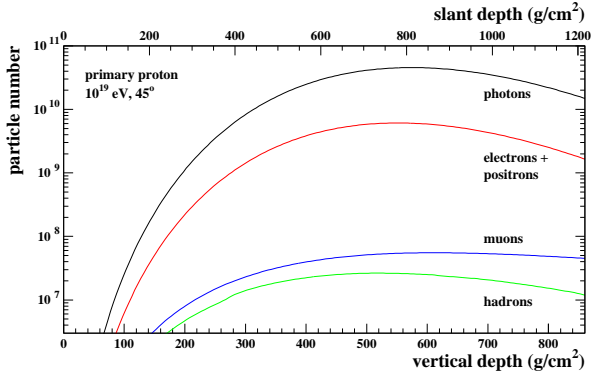}
\caption{Particle distribution at ground from a shower of 10 EeV at
an angle of 45$\circ$.}
\label{rcsfig06}
\end{figure}

We show in the Figure \ref{rcsfig06} a typical particle distribution
at the ground, for a shower with an energy of 10 EeV, with a zenith
angle of 45$^\circ$. Although gammas and electrons and positrons
dominate the shower, at large distances from the core and at large
angles, muons come to dominate the particle distributions.

\section{Auger data}

\begin{figure}[htb]
\centering\includegraphics[width=8.4cm]{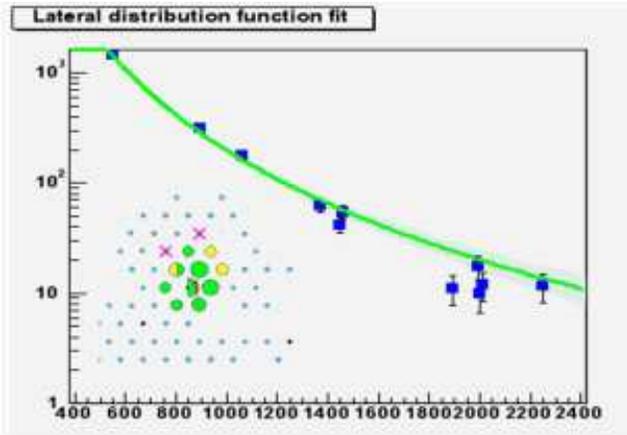}
\caption{A typical event registered by the SD system, shown in the
Event Display. The shower has a zenith angle of 30$^\circ$ and 10
stations are used in the fitting of the lateral distribution
function. The signal is quoted in VEM (Vertical Equivalent Muon).}
\label{rcsfig07}
\end{figure}

Since the begining of operations in 2001, Auger has been taking data
continuously, first in the Engineering Array mode, which is already
reported \cite{aug04}.  Since January 2004 it is taking data in
Physics Mode, with results that will be reported in the following. A
typical example of an event recorded by the SD system is shown in
the Figure \ref{rcsfig07}. The energy of the shower is infered from
the lateral distribution function ({\sc ldf}), fitted from density
of particles hitting each tank at distinct distances \cite{hil71,
dai88}. The density of particles at 1~km from the core, S(1000), is
quite independent of the nature of the primary cosmic ray, according
to distributions extracted from shower simulation programs, as
mentioned before. Each station in the array is continuously
calibrated by collecting at regular intervals signals generated by
single muons. From this signal the value of a VEM (vertical
equivalent muon) is extracted, taking into account variations in the
atmospheric temperature and pressure.

\begin{figure}[htbp]
\centering\includegraphics[width=8.5cm]{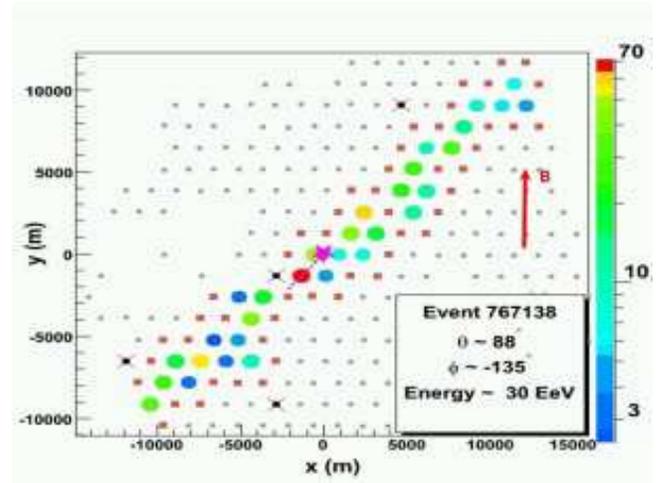}
\caption{Event with a large zenith angle, 88$^\circ$ in this
case. Most of the signal is carried by the surviving muons, once a
large part of the electromagnetic shower has been dissipated.}
\label{rcsfig08}
\end{figure}

The Auger ground stations are sensitive to very inclined showers
once it has a reasonable cross-section due to the 1.2~m column of
water. The very inclined showers will have transversed a larger
amount of atmospheric matter before hitting the station and a large
part of the electromagnetic shower will have dissipated, allowing
for a much larger relative muon component. An example of this class
of event is displyed in Figure \ref{rcsfig08} with a shower hitting
31 stations, coming with a zenith inclination of 88$^\circ$.

\begin{figure}[tbhp]
\centering\includegraphics[width=8.4cm]{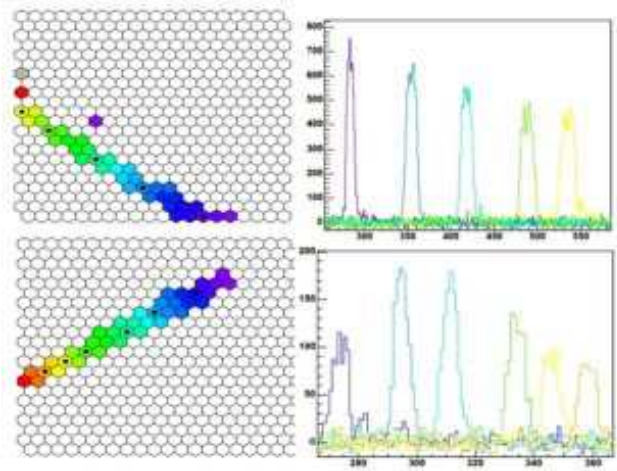}
\caption{Typical FD events. The camera display represent
the shower projected on the sky. The color code just identify the
pixel. The black dots on the pixels mark those exhibited on the
right. The time axis on the right side histogram is set in units of
100 ns, the label 350 means 35 $\mu$s. Top: Typical Laser shot
event, where the light goes from the ground up. Lower: Regular
event.}
\label{rcsfig09}
\end{figure}

The structure of the FD events is exhibited in Figure
\ref{rcsfig09}, taken from the event monitoring system. On both
pictures the map of the pixels are shown as if one would be facing
the sky. So, the first event, a typical low energy shower, is flying
top-down, right to left. The histogram on the top-right of the
display shows the time structure of the signal in a selection of the
pixels. The colour code is there just to correlate the pixel and the
signal. Each bin on the right plot corresponds to 100 ns, so that
the span of the signal from 260 to 360 units, actually corresponds
to a time interval of 10$\mu$s. The bottom picture of Figure
\ref{rcsfig09} represents a typical event propagating downwards,
while the top part there is an laser shot event, with a time
structure going upwards.

\begin{figure}[tbh]
\centering\includegraphics[width=8.4cm]{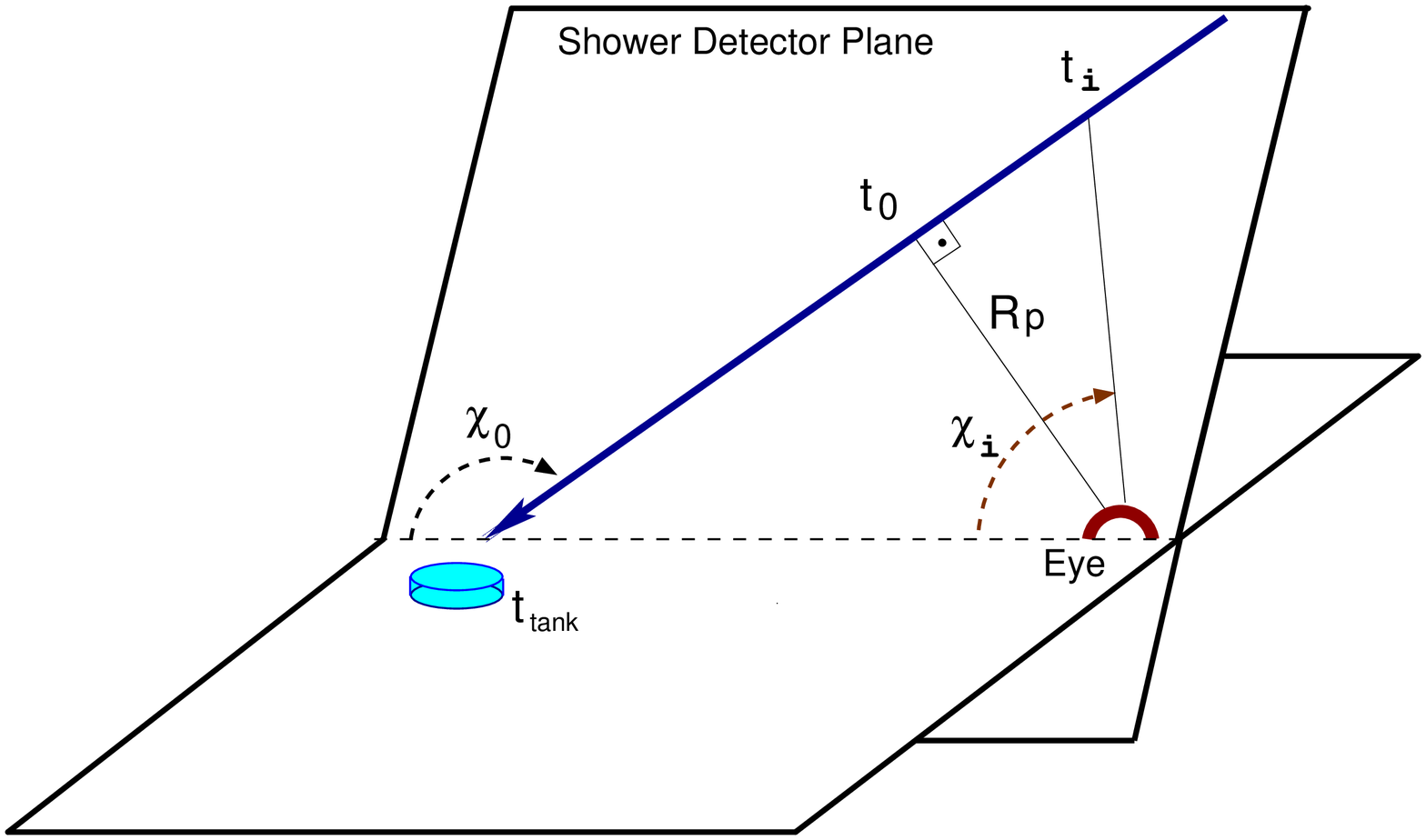}
\caption{Shower detector plane.}
\label{rcsfig10}
\end{figure}

\begin{figure}[tbhp]
\centering\includegraphics[width=8.9cm]{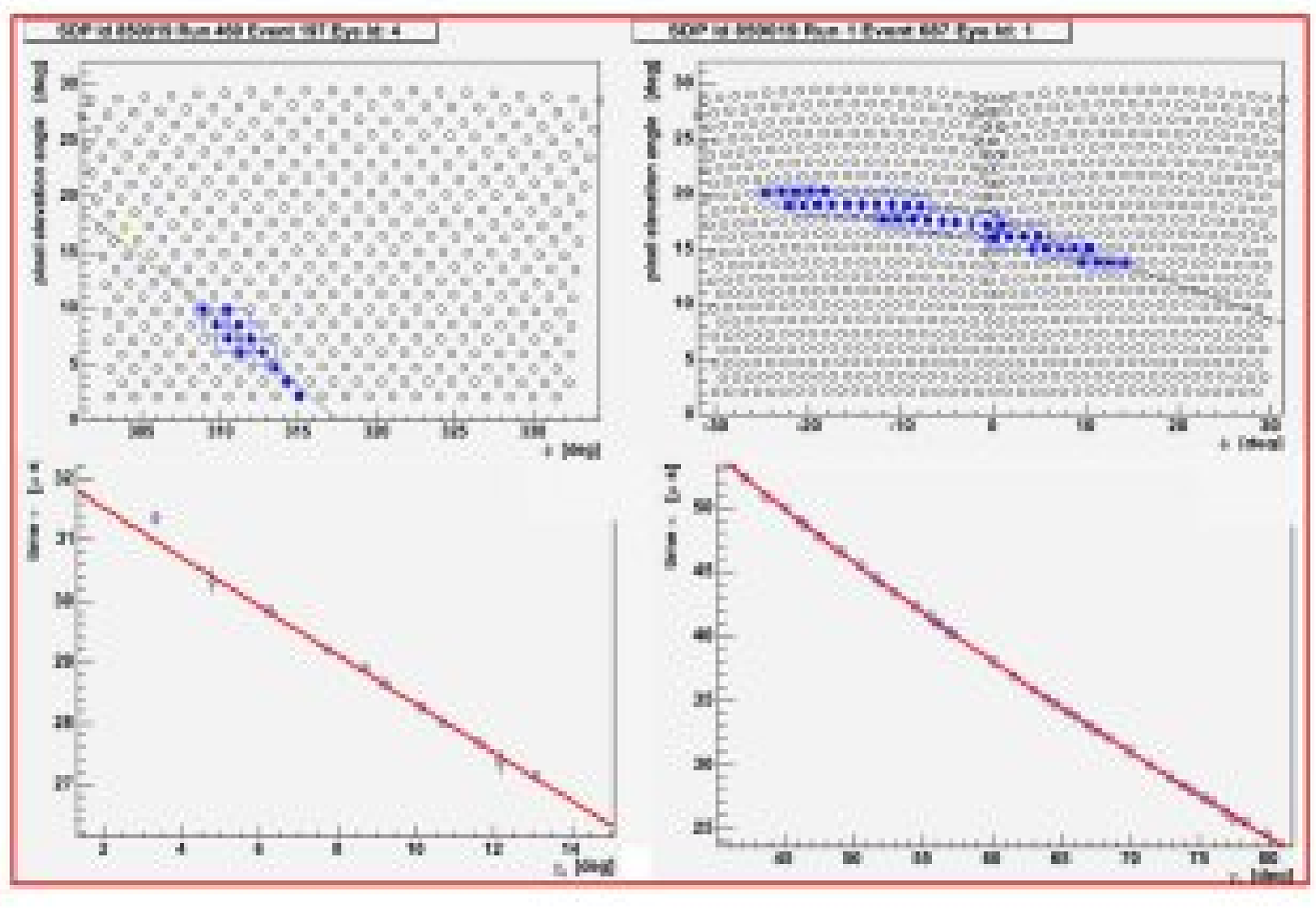}
\centering\includegraphics[width=8.4cm]{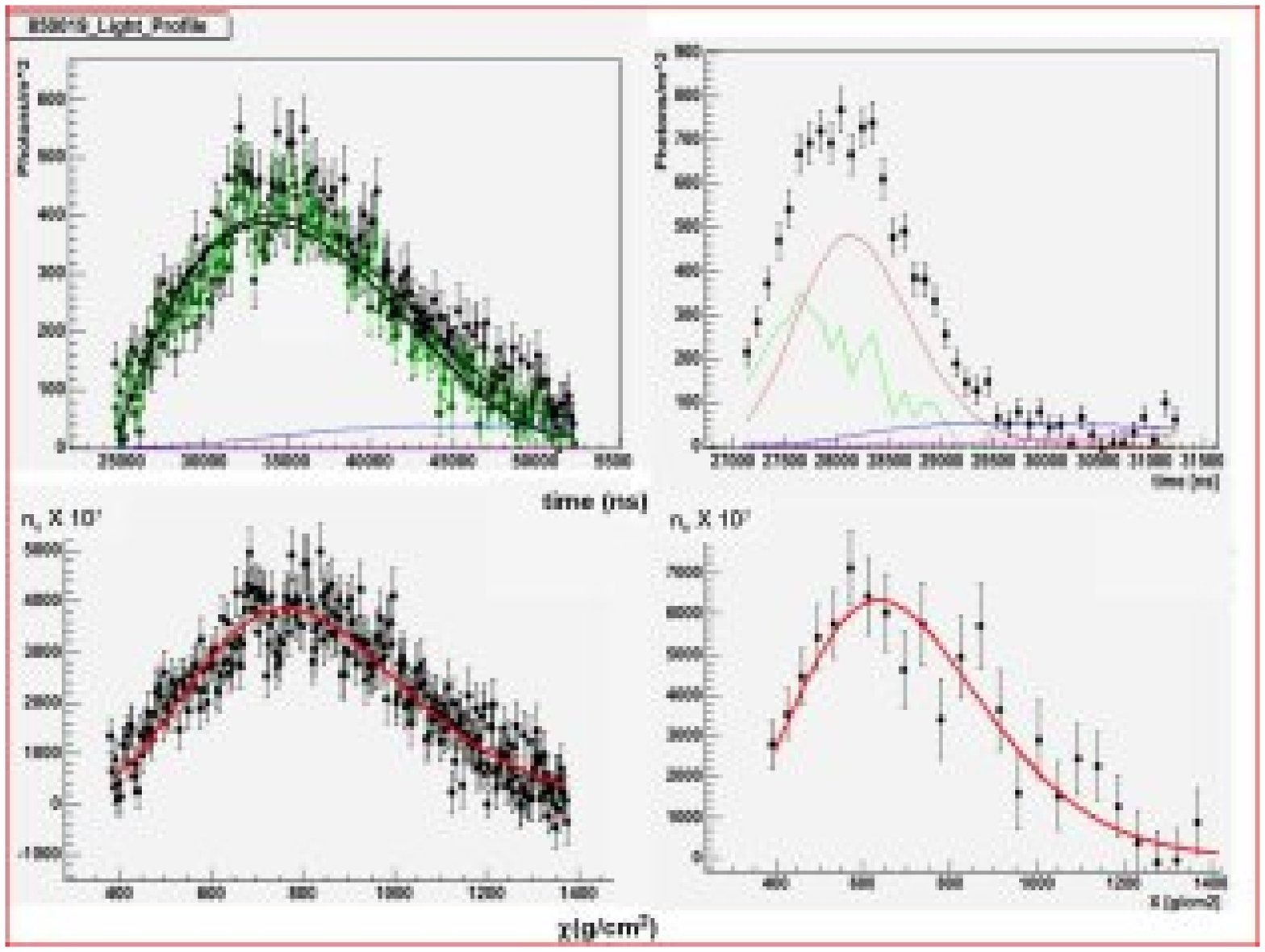}
\caption{Event seen in stereo mode. The left hand side is the view
from Los Leones and the right from Coihueco. The lower plots shows
the time {\it versus} angle correlation as seen by each eye. The fit
parameters are shown in the plots. Lower: The time profile and the
longitudinal profile of the event above}
\label{rcsfig11}
\end{figure}

To reconstruct the geometry of the shower, first the shower detector
plane ({\sc sdp}) is infered by optimizing the line of light
crossing the camera, using the signals as weight. The best estimate
of the normal vector to the {\sc sdp}, $\vec{n}_{SDP}$, is obtained
by minimizing
\[
\chi^2 = \sum_i w_i[\vec{n}_{SDP}\cdot \vec{r}_i]^2
\]
where the signal measured in pixel $i$ is used with the weight $w_i$
and the $\vec{r}_i$ corresponds to the direction pointing to the
source in the sky. The three dimensional geometry is recovered using
the angular velocity of the signal. For each shower pixel $i$ the
average time of the arrival of the light at that pixel field of
view, $t_i$, is determined from the FADC traces. The expression
\cite{som95},
\[
t_i =
t_0 + \frac{R_p}{c}\tan\left[\frac{(\chi_0 - \chi_i)}{2}\right],
\]
allow for a fitting of the shower parameters, $R_p$, $\chi_0$ and
$t_0$. Here, $c$ is the velocity of light, $R_p$ the shower distance
of closest approach to the detector and $t_0$ the time at which the
shower point reaches the position of closest approach. $\chi_i$,
indicated on Figure \ref{rcsfig10}, is the direction of the pixel
$i$ projected onto the {\sc sdp} and $\chi_0$ is the angle between
the shower axis and the direction from the detector to the shower
landing point. Figure \ref{rcsfig11} shows an example  of time {\it
versus} angle plot for a shower seen in stereo mode. However, this
procedure is not free of ambiguities, which can be resolved with the
input from the SD system. The timing information and location from a
station closest to the shower landing point can be related to the
time $t_0$,
\[
t_0 = t_{tank} - \frac{\vec{R}_{tank}\cdot\vec{S}_{shw}}{c},
\]
where $\vec{R}_{tank}$ is the vector connecting the fluorescence
detector to the ground station and $\vec{S}_{shw}$ is the unit
vector associated to the shower propagation.

To measure the energy, the light emitted by the source is
reconstructed making the corrections for the atmosphere attenuation
and than subtracting the Cherenkov component of the signal,
identifying the fluorescence component. The time profile and the
longitudinal profile for a stereo event event is exhibited  in
Figure \ref{rcsfig11} for both views of the event. The line fitting
the longitudinal profile represents the Gaisser-Hillas function
\cite{gai77}.

Although there is much work to do to improve the quality of the
measurements done with the SD and FD components, there are some
preliminary tests that points to overall quality of the data.
In particular, the correlation in the estimation of the energy
of hybrid events, measured by the SD and the FD, is quite consistent.

\section{Auger results}

\subsection{A first estimate of the cosmic ray spectrum above 3 EeV}

For the first attempt to estimate the cosmic ray spectrum above 3
EeV the analysis was based on a selection of events collected from
January 1st, 2004, through June 5th, 2005. The event acceptance
criteria and exposure calculation are described in a separate paper
\cite{Allard:2005he}. Events included in the sample have zenith
angles of less than 60$^\circ$ and energy greater than  3 EeV,
resulting in a selection of 3525 events. The array is fully
efficient for detecting such showers, so the acceptance at any time
is given by the simple geometric aperture. The cumulative exposure
adds up to 1750 km$^2$ sr yr, which is about 7\% greater than the
total exposure obtained by AGASA \cite{tak03}, along the years. The
average array size during the time of this exposure was 22\% of what
will be available when the southern site of the Observatory has been
completed.

Assigning energies to the SD event set is a two-step process, of
which the first is to assign an energy parameter $S_{38}$ to each
event. Then the hybrid events are used to establish the rule for
converting S38 to energy. The energy parameter $S_{38}$ for each
shower comes from its experimentally measured $S(1000)$, which is
the time-integrated water Cherenkov signal that would be measured by
a tank 1000 meters from the core.

The signal $S(1000)$ is attenuated at large slant depths. Its
dependence on zenith angle is derived empirically by exploiting the
nearly isotropic intensity of cosmic rays. By fixing a specific
intensity $I_0$ (counts per unit of $\sin^2 \theta$), one finds for
each zenith angle the value of $S(1000)$ such that $I(> S(1000)) =
I_0$. We calculated a particular constant intensity cut curve
CIC($\theta$) relative to the value at the median zenith angle
($\theta \approx 38^\circ$). Given $S(1000)$ and $\theta$ for any
measured shower, the energy parameter $S_{38}$ is defined by $S_{38}
\equiv S(1000)/CIC(\theta)$. It may be regarded as the $S(1000)$
measurement the shower would have produced if it had arrived
38$^\circ$ from the zenith.

\begin{figure}[htbp]
\centerline{\includegraphics[width=8.4cm]{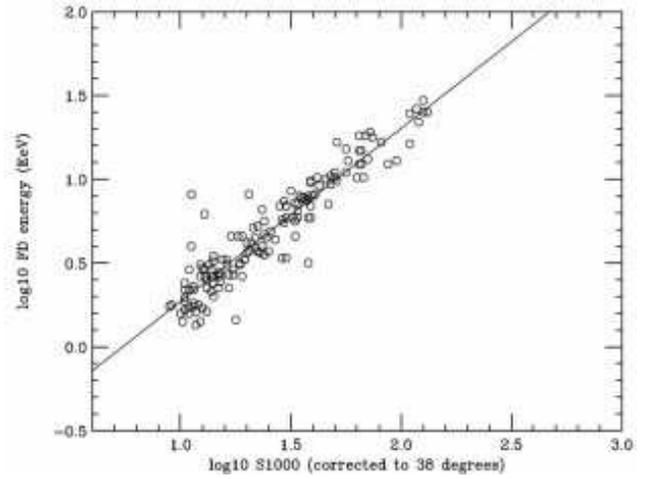}}
\caption{Shower energy for hybrid events as a function of the surface detector parameter
$S_{38}$.}
\label{rcsfig12}
\end{figure}

The $S_{38}$ parameter is well correlated to the energy extracted
from well reconstructed hybrid showers, as can be seen at Figure
\ref{rcsfig12}. The fitted line gives an empirical rule for
assigning energies in EeV, based on the $S_{38}$ parameter, given in
VEM,
\[
E = 0.16 \times S^{1.06}_{38} = 0.16 \times \left[S(1000)/CIC(\theta)\right]^{1.06}
\]

\begin{figure}[htbp]
\centerline{\includegraphics[width=8.4cm]{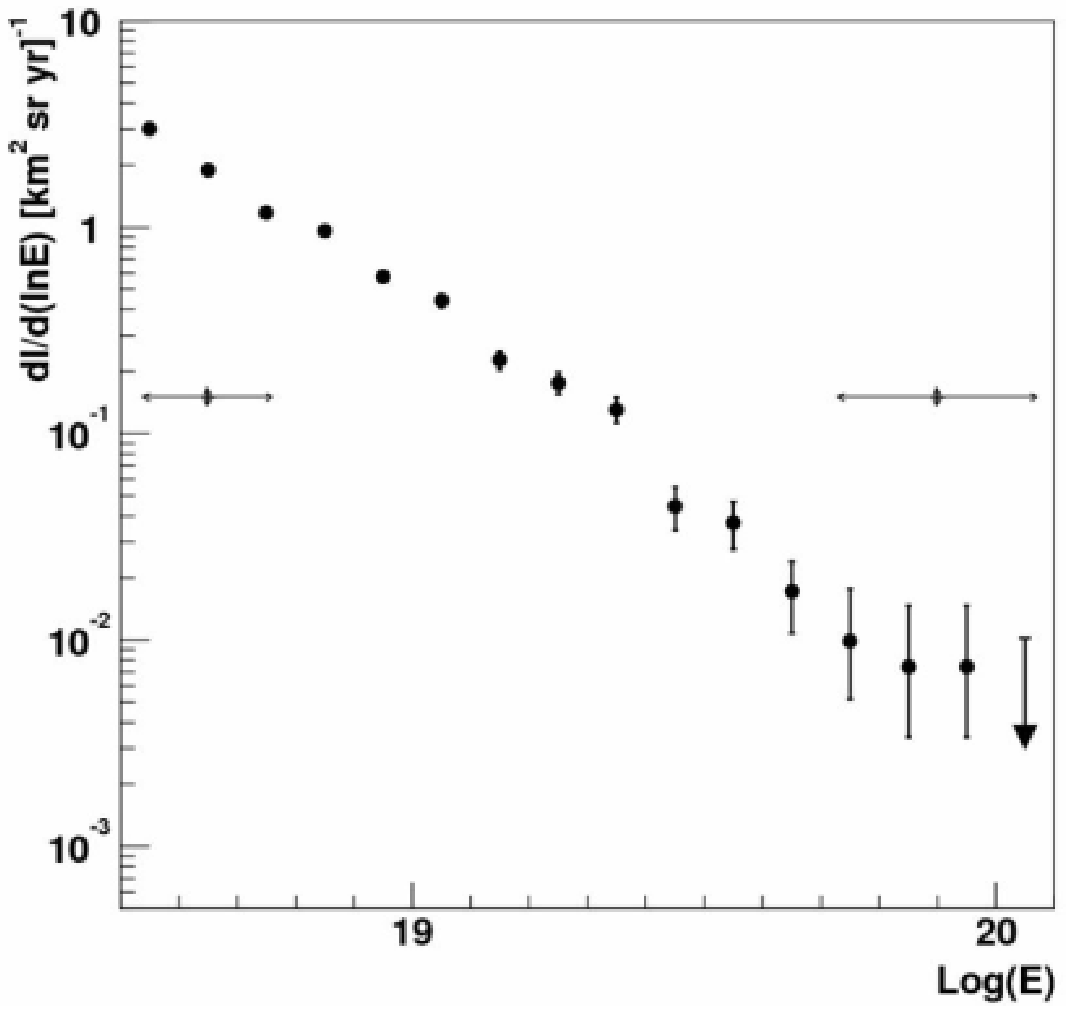}}
\centerline{\includegraphics[width=8.4cm]{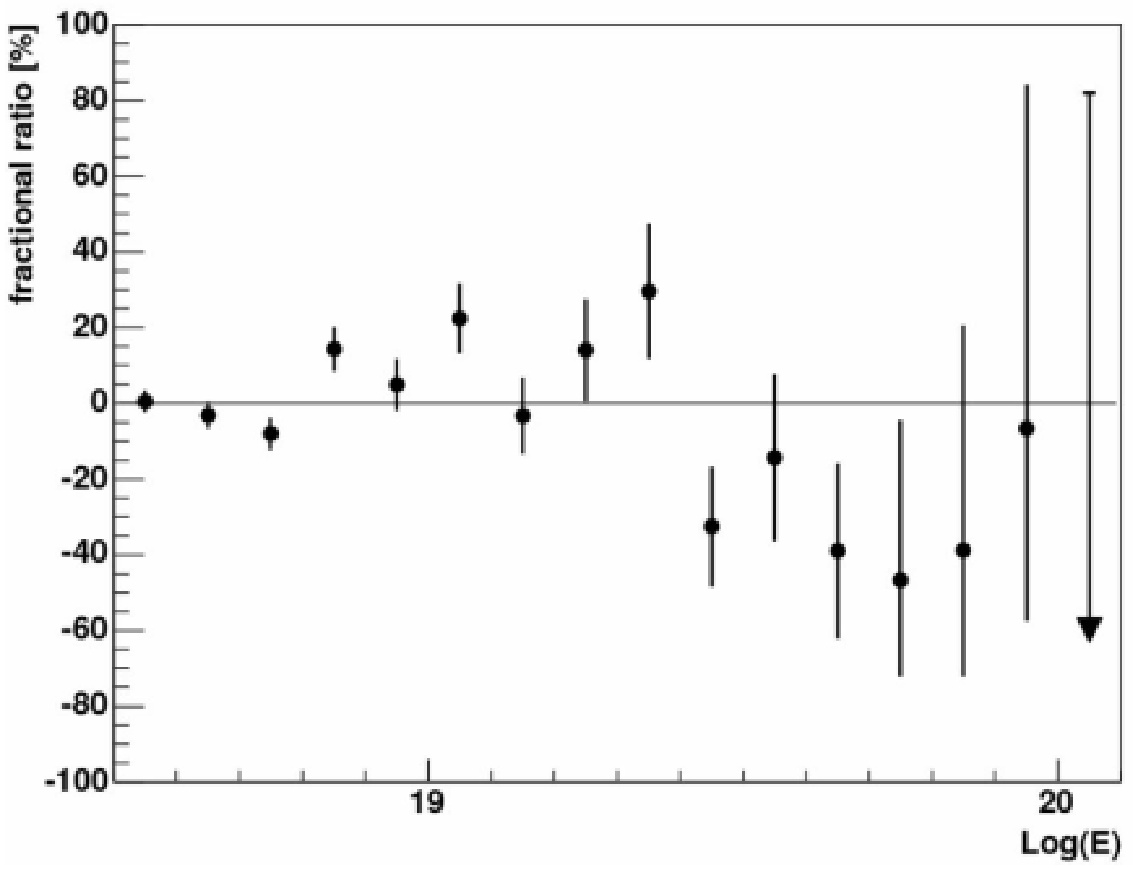}}
\caption{Top: Estimated spectrum. Lower: Deviation of the behaviour of the spectrum,
in relation to a power law (see text).}
\label{rcsfig13}
\end{figure}

On the top picture of  Figure \ref{rcsfig13} we show the preliminary spectrum extracted
from the data up to June 2005. Plotted on the vertical axis is the
differential intensity
\[ \frac{dI}{d\ln E} \equiv E \frac{dI}{dE}.
\]
Error bars on the points indicate the statistical uncertainty (or 95\% CL
upper limit). Systematic uncertainty is indicated by double arrows
at two different energies. On the bottom of Figure \ref{rcsfig13}, the percentage
deviation from the best-fit power law, $100 \times ((dI/d(\ln E)-F)/F$, is shown.
The fitted function is
\[
F = 30.9 \pm 1.7 \times {\rm (E/EeV)}^{-1.84 \pm 0.03},
\]
with a chisquare per degree of freedom of 2.4.

\section{Anisotropy studies around the Galactic Center}

The AGASA experiment \cite{Teshima:1999xy, hay99} and the SUGAR
experiment \cite{Bellido:2000tr} have reported an excess of cosmic
rays coming from the general direction of the center of our galaxy,
with energies in the EeV range. There is a very massive black hole
at the center of the galaxy (CG), which provides a natural candidate
for a cosmic ray accelerator to very high energies. Thus this region
provides an attractive target for anisotropy studies with the Pierre
Auger Observatory.

For this analysis a data set was selected comprising all the SD
events collected from January 1st, 2004, until June 6th, 2005, with a
zenith angle less than 60$^\circ$. Events from the Surface Detector
that passed the 3-fold or the 4-fold data acquisition triggers and
satisfyed our high level physics trigger (T4) and our quality
trigger (T5) \cite{Allard:2005vk} were selected. The T5 selection is
independent of energy and ensures a better quality for the event
reconstruction. This data set has an angular resolution better than
2.2$^\circ$ \cite{Bonifazi:2005ns} for all of the 3-fold events,
regardless of the zenith angle considered,  and better than
1.7$^\circ$ for events with multiplicities larger than 3 SD
stations.

\begin{figure}[htbp]
\centerline{\includegraphics[width=8.4cm]{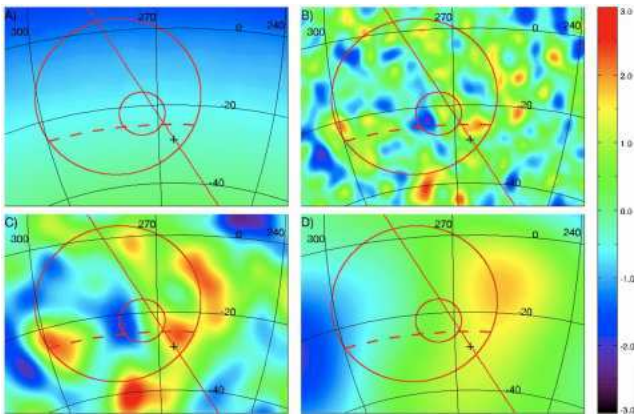}}
\caption{Lambert projections of the galactic centre region, GC
(cross), galactic plane (solid line), regions of excess of AGASA and
SUGAR (circles), AGASA f.o.v. limit (dashed line). The explanation
of each figure is detailed in the text.}
\label{rcsfig14}
\end{figure}

To estimate the coverage map, needed to construct excess probability
maps, a shuffling technique was used. Figure \ref{rcsfig14}A shows
the coverage map obtained from the SD sample in a region around the
GC. It has the same color scale as the significance maps, but
limited to the range 0 to 1. Figures \ref{rcsfig14}B, C and D,
present the chance probability distributions \cite{lima83} -- mapped
to positive Gaussian significance for excesses and negative for
deficits -- in the same region for various filtering and energy cuts
corresponding to the various searches. Figures \ref{rcsfig14}B is
the significance map in the range from 0.8 to 3.2 EeV, smoothed
using the individual pointing resolution of the events and a
1.5$^\circ$ filter (Auger like excess), while \ref{rcsfig14}C is the
same, smoothed at 3.7$^\circ$ (SUGAR like excess). Figure
\ref{rcsfig14}D covers the range from 1.0 to 2.5 EeV, smoothed at
13.3$^\circ$ (AGASA like excess). In these maps the chance
probability distributions are consistent with those expected as a
result of statistical fluctuations from an isotropic sky.

In the region where the AGASA excess was reported, there are 1155
events in the energy range from 1.0 to 2.5 EeV observed at the Auger
Observatory, which is to be compared to an expected sample of 1160.7
events, thus a ratio of $1.00 \pm 0.03$.

These results are not compatible with the excess reported by AGASA.
The level of 22\% excess that they observed would translate into a
$7.5\ \sigma$ excess. At the worst case scenario, where the source
would be made of nucleons and the bulk of the background at this
energy range made of much heavier nuclei (e.g. Iron), the
difference in detection efficiency of the Auger trigger at 1 EeV
would reduce the sensitivity to a source excess. However, using the
ratio of Fe (70 \%)  to proton (50 \%) efficiency at 1 EeV  (1.44,
an upper bound in the range from 1 to 2.5 EeV) a $5.2\ \sigma$ event
excess would still be expected from our data set.

We observe in the angular/energy window, where SUGAR claims an
excess, 144 events observed against 150.9 expected, with a ratio
$0.95 \pm 0.08$. With over an order of magnitude more statistics we
are not able to confirm this claim.

A search was performed for signals of a point-like source in the
direction of the galactic center using a 1.5$^\circ$ Gaussian filter
corresponding to the angular resolution of the SD
\cite{Bonifazi:2005ns}. In the energy range of 0.8 to 3.2 EeV, we
obtain 24.3 events observed, against 23.9 expected, with a ratio
$1.0 \pm 0.1$. A 95\% CL upper bound on the number of events coming
from a point source in that window is $n_s= 6.7$. This bound can be
translated into a flux upper limit ($\Phi_s$) integrated in this
energy range. In the simplest case in which the source has a
spectrum similar to the one of the overall CR spectrum
(${\rm d}N/{\rm d}E \propto E^{-3}$),
\[
\Phi_s = n_s\ \Phi_{CR}\ 4\ \pi\ \sigma^2/n_{exp},
\]
where $\sigma$ is the size of the Gaussian filter used.
Using
\[
\Phi_{CR} (E) = 1.5\ \xi\ \left(E/EeV\right)^{-3} \times 10^{-12}\
{\rm (EeV^{-1} m^{-2} s^{-1} sr^{-1})},
\]
where $\xi$, a number in the range from 1 to 2.5, denotes our
uncertainty on the CR flux ($\xi$ is around unity for Auger and 2.5
for AGASA), introducing the factor $\epsilon$ for the iron/proton detection
efficiency ratio ($1 < \epsilon < 1.6$ for E in the range from 0.8
to 3.2  EeV) and, integrating in that energy range we obtain :
\[
\Phi_s < 2.6\ \xi\ \epsilon\  \times 10^{-15} m^{-2} s^{-1} \ \ \ \ (95\%\ {\rm CL}).
\]

In the worst case scenario, where both $\xi$ and $\epsilon$ take their
maximum value, the bound is $\Phi_s = 10.6 \times 10^{-15}$ m$^{-2}$
s$^{-1}$, excluding the neutron source scenario, suggested
by the references \cite{hay99,bos03}, to account for the AGASA excess, or by
\cite{aha04a,aha04b}, in connection with the HESS measurements. More
details about the GC anisotropy studies with the Auger Observatory
data can be found in \cite{Abraham:2006ur}.

\section{An upper limit on the primary photon fraction}

The photon upper limit derived here is based on the direct
observation of the longitudinal air shower profile and makes use of
the hybrid detection technique and $X_{max}$ is used as discriminant
observable. The information from triggered surface detectors in
hybrid events considerably reduces the uncertainty in shower track
geometry.

The data were taken by the 12 fluorescence telescopes at the Los
Leones and Coihueco eyes, from January 2004 up to April 2005. The
number of deployed surface detector stations grew from  about 200 to
800 during this period. For the analysis, hybrid events were
selected, i.e. showers observed both, by  the surface tanks (at
least one) and the telescopes \cite{Mostafa:2005kd}. Even for one
triggered tank only, the additional timing constraint allows a
significantly improved geometry fit to the observed profile, which
leads to a reduced uncertainty in the reconstructed $X_{max}$.

\begin{figure}[htbp]
\centerline{\includegraphics[width=8.4cm]{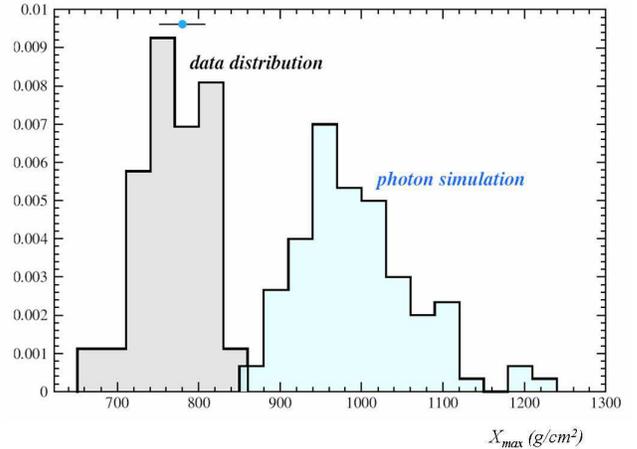}}
\caption{Example of $X_{max}$ measured in an individual shower of 16 EeV
(point with error bar) compared to the $X_{max}$ distribution expected
for photon showers (solid line). Also shown the $X_{max}$ distribution of
the data sample (dashed line).}
\label{rcsfig15}
\end{figure}

The reconstruction is based on an end-to-end calibration of the
fluorescence telescopes \cite{Bauleo:2005za}, on monitoring data of
local atmospheric conditions \cite{Cester:2005hk,Keilhauer:2005ja},
and includes an improved subtraction of Cherenkov light
\cite{Nerling:2005ew} and reconstruction of energy deposit profiles
for deriving the primary energy. In total, 16 events with energies
above $10^{19}$ eV are selected. The total uncertainty $\Delta
X^{tot}_{max}$ of the reconstructed depth of shower maximum is
composed of several contributions which, in general, vary from
event to event. A conservative estimate of the current $X_{max}$
uncertainties gives $\Delta X^{tot}_{max} \simeq 40$ g cm$^{-2}$.
Among the main contributions, each one in general well below
$\Delta X_{max}$ = 15 g cm$^{-2}$, are the statistical uncertainty
from the profile fit, the uncertainty in shower geometry, the
uncertainty in atmospheric conditions such as the air density
profile, and the uncertainty in the reconstructed primary energy,
which is taken as input for the primary photon simulation. For each
event, high-statistics shower simulations are performed for photons
for the specific event conditions. A simulation study of the
detector acceptance to photons and nuclear primaries has been
conducted. For the chosen cuts, the ratio of the acceptance to
photon-induced showers to that of nuclear primaries (proton or iron
nuclei) is $\epsilon = 0.88$. A corresponding correction is applied
to the derived photon limit. Figure \ref{rcsfig15} shows as an
example an event of 16 EeV primary energy observed with $X_{max}$ =
780 g cm$^{-2}$, compared to the corresponding $X_{max}$
distribution expected for simulated primary photons. With $< X^{\gamma}_{max}
>\ = 1020$ g cm$^{-2}$, photon showers are on average expected to
reach maximum at depths considerably greater than observed.
Shower-to-shower fluctuations are large due to the LPM effect (rms
of 80 g cm$^{-2}$) and well in excess of the measurement
uncertainty. For all 16 events, the observed $X_{max}$ is well below
the average value expected for photons. The $X_{max}$ distribution
of the data is also displayed in Figure \ref{rcsfig15}. More details
about this analysis can be found in \cite{Abraham:2006ar}.

\begin{figure}[htbp]
\centerline{\includegraphics[width=8.4cm]{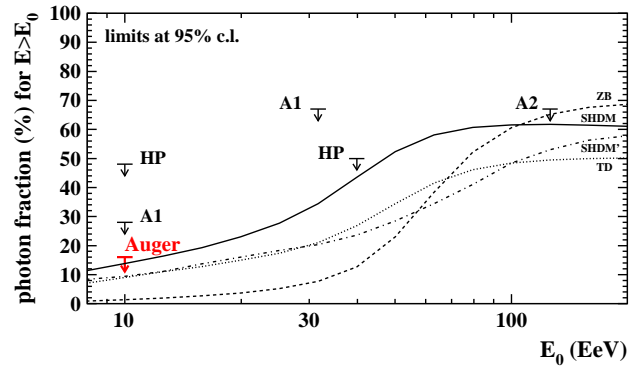}}
\caption{ Upper limits on cosmic-ray photon fraction compared with previous
experiments and estimates based on non-acceleration models (see main text).}
\label{rcsfig16}
\end{figure}

The statistical method for deriving an upper limit follows that
introduced in \cite{Risse:2005jr}. For the Auger data sample, an
upper limit on the photon fraction of 26\% at a confidence level of
95\% is derived. In Figure \ref{rcsfig16}, this upper limit (Auger)
is plotted together with previous experimental limits, from AGASA
(A1) \cite{Shinozaki:2002vv} , (A2) \cite{Risse:2005jr} and Haverah
Park (HP) \cite{Ave:2000nd} data and compared to some estimates
based on non-acceleration models (ZB, SHDM and TD from
\cite{Gelmini:2005wu} and SHDM' from \cite{Ellis:2005zz}). The
presented 26\% limit confirms and improves the existing limits above
10$^{19}$ eV. More details of this analysis can be found in
\cite{Abraham:2006ar} and references therein.

\section{Summary}

The Pierre Auger Observatory is still under construction but has
already the largest integrated exposure to high energy cosmic rays.
The combination of fluorescence and the surface detector
measurements allow for the reconstruction of the shower geometry and
its energy with much greater quality than what could be achieved
with either detector standing alone. Each of the detectors have
different systematics, allowing for valuable information for
cross-checking the results from each of them.

The observatory should be finished by mid 2007, accumulating by then
a much larger exposure than what was used for the preliminary
results presented here. That will allow for the search of
anisotropies in the southern sky, as well as the test of the
predicted GZK suppression.

\end{document}